# Combined Description of Pressure-Volume-Temperature and Dielectric Relaxation of Several Polymeric and Low-Molecular-Weight Organic Glass-Formers using "SL-TS2" Mean-Field Approach

Valeriy V. Ginzburg[1,*], Alessio Zaccone[2], and Riccardo Casalini[3]

[1]Department of Chemical Engineering and Materials Science, Michigan State University, East Lansing, MI 48824, USA

[2]University of Milan, Department of Physics, via Celoria 16, 20133 Milano, Italy

[3]Chemistry Division, Naval Research Laboratory, 4555 Overlook Avenue SW, Washington, USA

Corresponding author, email vvg851966@gmail.com

## Abstract

We apply our recently-developed mean-field "SL-TS2" (two-state Sanchez-Lacombe) model to simultaneously describe dielectric $\alpha$-relaxation time, $\tau_\alpha$, and pressure-volume-temperature (PVT) data in four polymers (polystyrene, poly(methylmethacrylate), poly(vinyl acetate) and poly(cyclohexane methyl acrylate)) and four organic molecular glass formers (ortho-terphenyl, glycerol, PCB-62, and PDE). Previously, it has been shown that for all eight materials, the Casalini-Roland thermodynamical scaling, $\tau_\alpha = f(Tv_{sp}^\gamma)$ (where $T$ is temperature and $v_{sp}$ is specific volume) is satisfied (Casalini, R.; Roland, C. M. *Phys. Rev. Lett.* **2014**, *113* (8), 85701). It has also been previously shown that the same scaling emerges naturally (for sufficiently low pressures) within the "SL-TS2" framework (Ginzburg, V. V. *Soft Matter* **2021**, *17*, 9094–9106.) Here, we fit the ambient pressure curves for the relaxation time and the specific volume as functions of temperature for the eight materials and observe a good agreement between theory and experiment. We then use the Casalini-Roland scaling to convert those results into "master curves", thus enabling predictions of relaxation times and specific volumes at elevated pressures. The proposed approach can be used to describe other glass-forming materials, both low-molecular-weight and polymeric.

Introduction

Glass transition is an important phenomenon impacting multiple properties of amorphous materials (mechanical, transport, thermophysical, etc.).[1–6] In particular, as a material approaches its glass transition from the higher-temperature, liquid state, its viscosity and characteristic relaxation time increase multiple orders of magnitude, while its coefficient of thermal expansion and heat capacity change abruptly (nearly discontinuously).[1,3,7–11] The glass transition temperature ($T_g$) depends on the molecular structure of the material and the applied pressure; it is also a weak function of the rate of temperature change (heating or cooling).

Over the past century, theoretical understanding of the glass transition has advanced significantly. The earliest models (Vogel,[12] Fulcher,[13] Tammann and Hesse[14]) described the viscosity or relaxation time as functions of temperature diverging at some finite temperature $T_0$ ($0 < T_0 < T_g$). The Vogel-Fulcher-Tammann-Hesse (VFTH) model – and the similar Williams-Landel-Ferry[15] (WLF) equation – was subsequently explained theoretically on the basis of "free volume" (Doolittle[16]) or "configurational entropy" (Adam and Gibbs [AG][17]). Other well-known empirical and semi-phenomenological models for the relaxation time as a function of temperature included MYEGA,[18] KSZ,[19] Avramov-Michev,[20] Elmatad-Chandler-Garraghan,[21] and others. Recently, Ginzburg[22,23] proposed a phenomenological "two-state, two-(time)scale" (TS2) model which postulated the existence of two separate "states", the low-temperature "Solid" and the high-temperature "Liquid", with both of them characterized by the Arrhenius dependence of the relaxation time on temperature.

The description of the relaxation time (or viscosity) should, ideally, be coupled to the description of the other properties, like specific volume or heat capacity. Here, we concentrate on specific volume. The dependence of specific volume (or density) on temperature and pressure is described by an equation of state (EoS). Those EoS (see, e.g., Rodgers[24]) include variety of empirical (such as the Tait equation[25,26]) and thermodynamic theory-based (Flory-Orwoll-Vrij,[27,28] Prigogine cell model,[29] Dee-Walsh modified cell model,[30,31] Simha-Somcynsky hole model,[32,33] Hartmann-Haque model,[34] Sanchez-Lacombe lattice fluid model,[35–37] and others) approaches. Then, one needs to estimate how the relaxation time couples to the density (or its fluctuations), using approaches like mode-coupling theory (MCT),[38] elastic cooperative nonlinear Langevin equation (ECNLE),[39–42] generalized entropy theory (GET),[43–47] cooperative free volume model,[48–52] "the shoving model",[53] and many others.

Even though there are multiple ways to construct a combined theory for the specific volume and relaxation time, it is important that such a theory satisfy a number of empirically observed constraints. In particular, the dielectric relaxation spectrum often is not a function of the temperature, T, and pressure, P, independently, but rather depends on some combined "state variable" X(T,P),[54] consistent, e.g., with the "coupling model" of Ngai and co-workers.[55] Casalini, Roland, and co-workers determined, in particular, that for many amorphous glass-formers, this constraint can be written in the form of the following scaling relationship, $\tau_\alpha = f(T\, v_{sp}^\gamma)$ (where T is temperature and $v_{sp}$ is specific volume).[56–61] Here, the specific volume is calculated based on the temperature and pressure using a parameterized EoS for the liquid state (above the glass transition). The parameter $\gamma$ is similar to the Grüneisen constant and can be interpreted within a revised entropy theory of glass transition.[56]

Very recently, Ginzburg[62,63] formulated a new semi-phenomenological model of glassy materials based on his earlier TS2 dynamic model and the two-state Sanchez-Lacombe EoS.[64,65] The combined model (labeled SL-TS2) was able to capture the Casalini-Roland scaling and correctly describe the $\alpha$-relaxation time and specific volume vs. temperature for two polymers, polystyrene (PS) and poly(methyl methacrylate) (PMMA). Here, we expand this analysis and apply the model to several other amorphous materials, both low-molecular-weight and polymeric.

## The Model and Experimental Methods

### The SL-TS2 Model

*Thermodynamics – the two-state Sanchez-Lacombe approach*

The SL-TS2 (two-state Sanchez-Lacombe) model was described in detail by Ginzburg;[62,63] here, we recap its main features. To begin with, we assume that the glass-forming amorphous material can be separated into small "clusters" or "cooperatively rearranging regions" (CRR). Each CRR might be a single molecule or polymer repeat unit, or a small group of molecules. It is further assumed that the material can exist in one of two states, "Solid" (S) or "Liquid" (L). The "solid" state is characterized by higher density (lower specific volume) and higher zero-temperature cohesive energy density (to be called simply "cohesive energy density" below), while the "liquid" state has lower density and lower cohesive energy density. Thus, "solid" and "liquid" states have different pressure-volume-temperature (PVT) dependences. Here, we model those dependences using Sanchez-Lacombe (SL) EoS, in which the finite compressibility of the material is modeled by allowing some "lattice sites" to be occupied by "voids" (white squares) with zero mass and zero cohesive energy. Furthermore, we assume that the "solid" CRR has lower volume, i.e., occupies fewer lattice sites as compared to the "liquid" CRR, at any given temperature. Thus, in Figure 1, the "solid" CRR (blue) occupies $r_S$ = 14 lattice sites, and the "liquid" CRR (red) occupies $r_L$ = 20 sites. Also, note that there are more voids in the "liquid" than in the "solid" because of the higher cohesive energy density in the latter. This approach is known as the two-state Sanchez-Lacombe or "Condo model".[64]

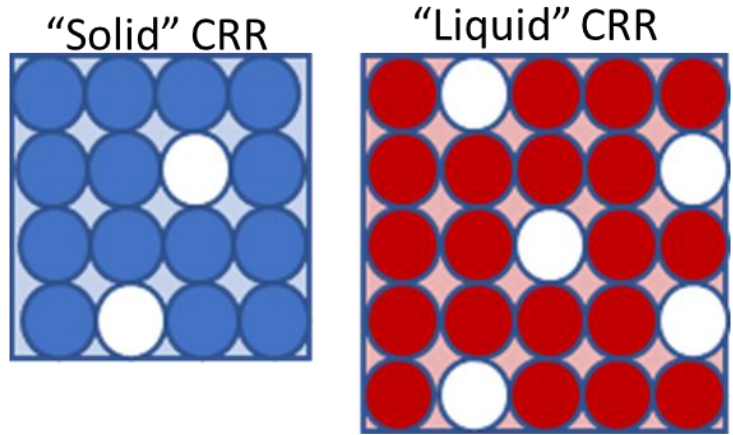

*Figure 1. SL-TS2 Model. The voids are shown in white, and the cell size is the same as the void size. In this example, $r_S$ = 14 (the number of non-void cells in the "solid" CRR), and $r_L$ = 20 (the number of non-void cells in the "liquid" CRR).*

The free energy of the material is given by,

$$G = k_B T \left[ \psi \ln \phi_S + (1-\psi) \ln \phi_L + \langle r \rangle \left\{ \frac{1}{v} - 1 \right\} \ln \phi_V \right]$$

$$- \varepsilon^* \frac{\langle r \rangle}{v} \left[ \phi_S^2 + 2\alpha_{LS} \phi_L \phi_S + \alpha_{LL} \phi_L^2 \right] + \frac{P v_0 \langle r \rangle}{v} \quad (1)$$

Here, $P$ is the pressure, $T$ is the absolute temperature, and $k_B$ is the Boltzmann's constant; $\psi$ is the mole or number fraction of CRRs in the "solid" state, and $v = 1 - \phi_V$ is the "occupancy".

Other variables in equation 1 are as follows: $\phi_L$, $\phi_S$, and $\phi_V$ are the volume fractions of "liquid", "solid", and void-occupied cells, respectively (to be defined below); $v_0$ is the volume of one cell (which itself is pressure-dependent). The lattice coordination number is $Z$, and the nearest-neighbor interaction energies are $\varepsilon_{SS}$, $\varepsilon_{SL}$, and $\varepsilon_{LL}$; the interaction energies including "void" cells are all set to zero, $\varepsilon_{SV} = \varepsilon_{LV} = \varepsilon_{VV} = 0$. We assume the Berthelot "geometric mean" rule, i.e., $\varepsilon_{SL} = (\varepsilon_{SS}\varepsilon_{LL})^{1/2}$. The number of cells occupied by a "liquid cluster" is $r_L$, and that of the "solid cluster" is $r_S$; obviously, $r_L > r_S$, so that the liquid density, $\rho_L = \dfrac{M}{r_L v_0}$, is smaller than the solid density, $\rho_S = \dfrac{M}{r_S v_0}$ (here, $M$ is the mass of the "cluster" or CRR, which is the material constant, independent of T and P). Also,

$$\langle r \rangle = r_S \psi + r_L (1-\psi) \tag{2a}$$

$$\phi_S = v \frac{\psi r_S}{\langle r \rangle} \tag{2b}$$

$$\phi_L = v \frac{(1-\psi) r_L}{\langle r \rangle} \tag{2c}$$

$$\phi_V = 1 - v \tag{2d}$$

$$\alpha_{LS} = \frac{\varepsilon_{LS}}{\varepsilon_{SS}} \tag{2e}$$

$$\alpha_{LL} = \frac{\varepsilon_{LL}}{\varepsilon_{SS}} \tag{2f}$$

$$\varepsilon^* = \frac{Z\varepsilon_{SS}}{2} \tag{2g}$$

To find thermodynamic equilibrium, we need to minimize G with respect to $\psi$ and $v$,

$$T\left[\ln\frac{\psi}{1-\psi} + \ln\frac{r_S}{r_L} + \frac{r_L - r_S}{\langle r \rangle} + (r_L - r_S)\left(\frac{1}{v} - 1\right)\ln\frac{1}{1-v}\right]$$

$$+ \frac{v}{\langle r \rangle^2}\left\{(r_L - r_S)\psi^2 r_S^2 - 2\psi r_S^2 r_L\right\}$$

$$+ 2\alpha_{LS}\frac{v}{\langle r \rangle^2}\left\{(r_L - r_S)\psi(1-\psi)r_S r_L - r_S r_L\left[(1-\psi)r_L - \psi r_S\right]\right\}$$

$$+ \alpha_{LL}\frac{v}{\langle r \rangle^2}\left\{(r_L - r_S)(1-\psi)^2 r_L^2 + 2(1-\psi)r_L^2 r_S\right\} - \frac{P}{v}(r_L - r_S) = 0$$

$$\tag{3a}$$

$$T\left[\ln(1-v) + v\left\{1 - \frac{1}{\langle r \rangle}\right\}\right] +$$

$$v^2\left[\left(\frac{\psi r_S}{\langle r \rangle}\right)^2 + 2\alpha_{LS}\left(\frac{\psi r_S}{\langle r \rangle}\right)\left(\frac{\{1-\psi\}r_L}{\langle r \rangle}\right) + \alpha_{LL}\left(\frac{\{1-\psi\}r_L}{\langle r \rangle}\right)^2\right] + P = 0$$

$$\tag{3b}$$

Here, $T = \frac{k_B T}{\varepsilon^*}$, and $P = \frac{Pv_0}{\varepsilon^*}$. We can re-write equations (3a) – (3b) in the following form,

$$\ln\frac{\psi}{1-\psi} + \Delta S - \frac{\Delta U + P\Delta V}{T} = 0 \tag{4a}$$

$$T\left[\ln(1-v)+v\left\{1-\frac{1}{\langle r \rangle}\right\}\right]+v^2 J+P=0 \tag{4b}$$

where,

$$\Delta S = \ln\frac{r_S}{r_L}+\frac{r_L-r_S}{\langle r \rangle}+(r_L-r_S)\left(\frac{1}{v}-1\right)\ln\frac{1}{1-v} \tag{5a}$$

$$-\Delta U = \frac{v}{\langle r \rangle^2}\left\{(r_L-r_S)\psi^2 r_S^2 - 2\psi r_S^2 r_L\right\}$$

$$+2\alpha_{LS}\frac{v}{\langle r \rangle^2}\left\{(r_L-r_S)\psi(1-\psi)r_S r_L - r_S r_L\left[(1-\psi)r_L - \psi r_S\right]\right\}$$

$$+\alpha_{LL}\frac{v}{\langle r \rangle^2}\left\{(r_L-r_S)(1-\psi)^2 r_L^2 + 2(1-\psi)r_L^2 r_S\right\}$$

$$\tag{5b}$$

$$\Delta V = \frac{(r_L-r_S)}{v} \tag{5c}$$

$$J = \left(\frac{\psi r_S}{\langle r \rangle}\right)^2 + 2\alpha_{LS}\left(\frac{\psi r_S}{\langle r \rangle}\right)\left(\frac{\{1-\psi\}r_L}{\langle r \rangle}\right) + \alpha_{LL}\left(\frac{\{1-\psi\}r_L}{\langle r \rangle}\right)^2 \tag{5d}$$

*Dielectric Relaxation – the TS2 Model*

The "two-state, two-(time)scale" (TS2) model has been proposed in ref.22 Within this approach, it is stipulated that the characteristic time for "jumps within the cage", $\tau_1$, is the same for both "L" and "S" elements and given by the Arrhenius temperature dependence with activation energy $E_L$. The characteristic times for "jumps into neighboring cages", on the other hand, is different for different elements – it is assumed to be equal to $\tau_1$ for jumps between adjacent "L" elements, and equal to a different, larger characteristic time $\tau_2$ for jumps involving at least one "S" element. The characteristic time $\tau_2$ is also postulated to have an Arrhenius temperature dependence with activation energy $E_S$; in general, $E_S > E_L$. The infinite-temperature limit for both $\tau_1$ and $\tau_2$ is assumed to be the same and labeled $\tau_\infty$. (Note that $\tau_\infty$, $E_S$, and $E_L$ are temperature-independent but pressure-dependent). As discussed in ref. 22, the dielectric $\alpha$- and $\beta$ (Johari-Goldstein)-relaxation times are given by,

$$\tau_\beta(T,P) \equiv \tau_1(T,P) = \tau_\infty \exp\left[\frac{E_L(P)}{RT}\right]$$

(6a)

$$\tau_\alpha(T,P) = \left(\tau_2(T,P)\right)^{\psi(T,P)} \left(\tau_1(T,P)\right)^{1-\psi(T,P)}$$

$$= \tau_\infty \exp\left[\frac{E_L(P)}{RT} + \frac{E_S(P) - E_L(P)}{RT}\psi(T,P)\right]$$

(6b)

Equation (6a) is based on the assumption that the Johari-Goldstein relaxation time is comparable or roughly equal to the "liquid" characteristic time $\tau_1$. Equation (6b) is an estimate of the $\alpha$-relaxation time (the time for a particle to diffuse out of its cage) in a mixture of L and S elements based on the effective medium approach (see Supporting Information to ref. 23) The solid fraction, $\psi$, obviously depends on the temperature and pressure; however, it can also depend on the cooling rate or other factors in the non-equilibrium case, as will be discussed below.

As we pointed out in earlier papers,[22,23,62] the $\alpha$-relaxation as a function of inverse temperature (Arrhenius analysis) can be visualized as a sigmoidal function, with lower slope corresponding to the "L" state, higher slope corresponding to the "S" state, and an abrupt transition between the two asymptotic regions. The center of this transition corresponds to the point where $\psi = 0.5$ (equal balance between "S" and "L"); we denote this temperature as $T_X(P)$. Note that, unlike $T_g$, $T_X$ is an equilibrium quantity, independent of cooling rate or other experimental conditions.

*Equations of evolution for non-equilibrium case (isobaric cooling from equilibrium melt)*

To describe the non-equilibrium behavior of $v$ and $\psi$ during, e.g., cooling from high-temperature equilibrium phase, we used a simple "relaxation time approximation", stipulating[62] that the relaxation time for $\psi$ is the JG $\beta$-relaxation and the relaxation time for $v$ is the (often much slower) $\alpha$-relaxation,

$$\frac{d\psi}{dt} = \frac{\psi^* - \psi}{\tau_\beta} \tag{7a}$$

$$\frac{dv}{dt} = \frac{v^* - v}{\tau_\alpha} \tag{7b}$$

Here, we define $\psi^*$ and $v^*$ in the following way. First, we solve equations (3a)—(3b) (or (4a) – (4b)) to obtain the equilibrium values, $\psi_{eq}(T,P)$ and $v_{eq}(T,P)$. Next, we set $v^* = v_{eq}(T,P)$ and update $v$ using equation (7b). Finally, we re-calculate $\psi^*$ by solving equation (3a) with the new value of $v$. Note that this approach is similar in spirit to well-known Tool-Narayanaswami-Moynihan (TNM)[66–68] and Kovacs–Aklonis–Hutchinson–Ramos (KAHR)[69] models.

To simplify the non-equilibrium modeling even further, we consider the following assumption. Let us define $T_g$ ($\geq T_X$) as the temperature below which the temperature change becomes faster than the $\alpha$-relaxation, i.e., $T_g^{-1}|q|\tau_\alpha(T_g) \cong 1$ (where $q = \dfrac{dT}{dt}$ is the cooling rate). In this case, for $T > T_g$, both $\psi$ and $v$ equilibrate fully, while for $T < T_g$, $v = v_{eq}(T_g)$ does not change, while $\psi = \psi^*(T;v)$ still continues to increase as the temperature is decreased, but significantly slower than in the equilibrium limit. For more details, see ref.[63]

*The Relaxation-Temperature-Specific Volume ($\tau$TV) Scaling*

The specific volume of a material can be expressed in terms of $v$ and $\psi$ as,

$$v_{sp} = v_{sp,0} \frac{\langle r \rangle}{r_S v} = v_{sp,0} \frac{r_S \psi + r_L(1-\psi)}{r_S v} \tag{8}$$

where $v_{sp,0}$ can depend on pressure but not on temperature. The scaling relationships proposed by Casalini, Roland, and co-workers suggest that,[56–58]

$$\tau_\alpha(T,P) = f\left[T\left(v_{sp}(T,P)\right)^\gamma\right] \tag{9}$$

where $\gamma$ is a material-dependent constant, independent of $T$ and $P$. This scaling means that the state of the material depends not on $T$ and $P$ independently, but on a combined state variable $X = T\left(v_{sp}(T,P)\right)^\gamma$. On the other hand, in refs.62,63 we demonstrated that in the limit of sufficiently low pressures, equations 6a and 6b can be re-written as,

$$\tau_\beta(T,P) = \tau_\infty \exp\left[\frac{E_L(P)}{RT_X(P)}(1+Z)\right] \tag{10a}$$

$$\tau_\alpha(T,P) = \tau_\infty \exp\left[\frac{E_L(P)}{RT} + \frac{E_S(P)-E_L(P)}{RT}\psi(T,P)\right]$$

$$= \tau_\infty \exp\left[\frac{E_L(P)}{RT_X(P)}(1+Z) + \frac{E_S(P)-E_L(P)}{RT_X(P)}(1+Z)\psi(Z)\right]$$

(10b)

Here, $Z = \frac{T}{T_X(P)} - 1$ is a different "state variable" that also combines the effects of temperature and pressure. As discussed in ref. 62, the solid fraction $\psi(T,P) \equiv \psi(Z)$ for the equilibrium case (above the glass transition); also, for constant-cooling-rate experiments performed at different pressures, a similar relationship holds as well, $\psi(T,P;q) \equiv \psi\left(\frac{T}{T_g(P)};q\right)$ (the ratio $T_g/T_X$ is slightly cooling-rate dependent, but pressure-independent for any given cooling rate $q$).

Going back to the (τTV) scaling relationships, we can substitute equation (8) into equation (9) to obtain,

$$\tau_\alpha(T,P) = f\left[T(v_{sp}(T,P))^\gamma\right] = f\left[\frac{T_X(P)}{1+Z}\left(v_{sp,0}(P)\frac{r_S\psi(Z)+r_L\{1-\psi(Z)\}}{r_S v(Z)}\right)^\gamma\right]$$

(11)

To satisfy the scaling, we need to eliminate the explicit pressure dependence in equation 11. This is accomplished if one assumes that all the temperature- and energy-related model parameters scale with pressure as,

$$Y(P) = Y(0)\exp\left(\frac{P}{P_{0,T}}\right) \quad (12a)$$

(here, $Y = T_x, T_g, \varepsilon_{ij}, E_L,$ and $E_S$), while the specific volume scales as,

$$v_{sp0}(P) = v_{sp0}(0)\exp\left(-\frac{P}{P_{0,v}}\right) \quad (12b)$$

Here, $P_{0,v} \equiv P^* = B$ is the zero-temperature bulk modulus, and $P_{0,T} = P_{0,v}/\gamma$. This analysis provides an alternative description of the Casalini-Roland parameter $\gamma$ as the ratio of the bulk modulus $B$ and the inverse of the slope of the dependence of $ln(T_g)$ on $P$. For more discussion, see ref.[62] Note that this scaling (equations 12a and 12b) holds only for relatively small pressures, $P \ll P_{0,T}$; for larger pressures, we expect the specific volume to saturate at some finite value rather than decaying all the way to zero. As will be shown below, $P_{0,T} \geq 1,000$ *MPa*, so the SL-TS2 scaling relationships should hold fairly well for pressures up to at least 200 – 300 *MPa*. It is important to note that the Casalini-Roland scaling can extend to significantly larger pressures where equations (12a) and (12b) are no longer valid and other mechanisms might apply. Another important difference is that the scaling relationships of equations (12a) and (12b) provide a way to scale dynamics and specific volume in the glassy state, while the Casalini-Roland scaling has been verified only in the liquid state.

*PVT Measurements*

In this paper we utilized the equation of state (EoS) data for eight liquids and polymers: glycerol,[70,71] phenylphthalein-dimethylether (PDE),[72] polychlorinated biphenyls (PCB),[73] o-terphenil (OTP),[59] polycyclohexylmethacrylate (PCHMA),[74] polyvinylacetate (PVAc),[75,76] polystyrene (PS),[77–79] polymethylmetacrylate (PMMA).[80]

The data reported in this paper have been previously published with a detailed description of the experimental setup. In general, to determine the EoS for glass forming liquids, volume changes are measured as a function of pressure and temperature using a set up like the Gnomix instrument (see, e.g., Zoller and Walsh[81]). A liquid (usually mercury because of its large density) serve as the confining medium, in order to maintain hydrostatic conditions when the sample solidifies by crystallization or vitrification. The differential data are then converted to specific volumes, $v$, using the value determined at ambient conditions using the buoyancy method (Archimedes' principle) or helium picnometer.

*Dielectric Relaxation Measurements*

The relaxation data were taken from earlier publications, as follows: glycerol,[70,71] PDE,[82] PCB,[73] OTP,[59,83,84] PCHMA,[74] PVAc,[75,85] PS,[77] PMMA;[80,86] we refer the readers to those publications for experimental details. In general, typical dielectric measurements are carried out using a parallel plate geometry and the dielectric constant is extracted from impedance measurements performed using impedance analyzers over a broad band of frequency. For measurements at elevated pressure, the sample cell is immersed in a pressure transmitting fluid and contained in a thermalized high pressure Manganin cell, with pressure controlled using hydraulic pumps. Proper choices to isolate the sample cell from the pressure transmitting fluid with small dielectric losses is necessary for accurate measurements.

## Results and Discussion

### *The Relaxation Time-Density-Temperature Scaling*

Before describing our results in detail, we highlight our approach to the "τTV" scaling. To do this, we use the example relaxation and specific volume data for PDE. Let us consider the relaxation data (Figure 2). The data are collected at various temperatures as a function of the pressure. We collapse all the curves onto a single master curve by introducing the following "effective temperature", $T_{eff}$,

$$T_{eff}(T, P) = T \exp\left(-\frac{P}{P_{0,T}}\right) \qquad (13)$$

Thus, $T_{eff} = T$ for $P = 0$ (ambient pressure).

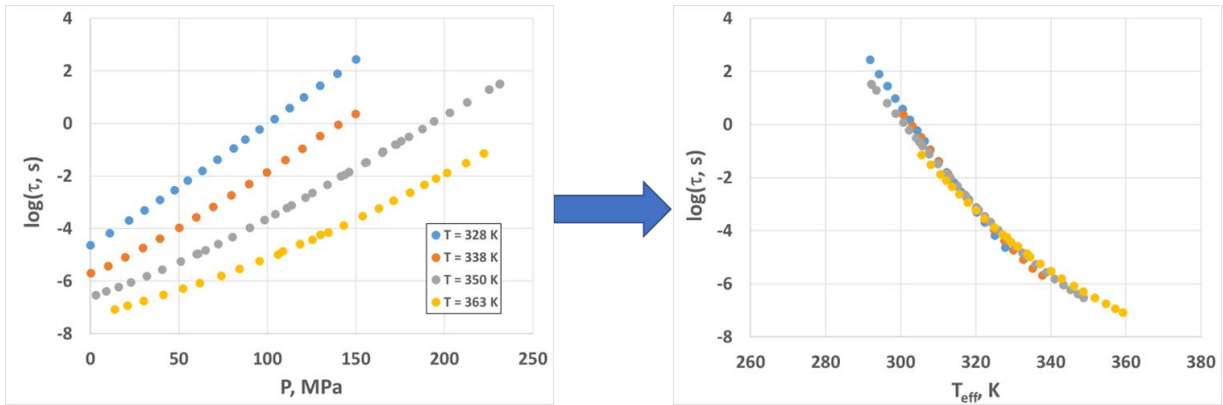

*Figure 2. Relaxation time scaling illustration based on PDE data. Left panel – dielectric relaxation time vs. pressure measured for several temperatures. Right panel – same data where the relaxation time is plotted against the effective temperature calculated using equation 13.*

The specific volume data are analyzed in a similar fashion (Figure 3). The temperature axis is scaled according to equation 13, while the specific volume axis is rescaled in a similar fashion,

$$V_{eff}(T,P) = v_{sp}(T,P)\exp\left(\frac{P}{P_{0,V}}\right) \quad (14)$$

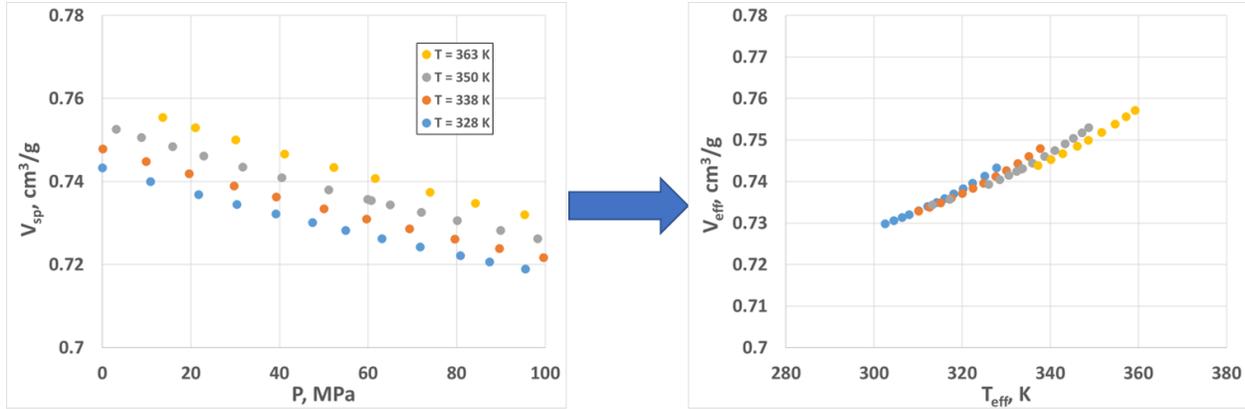

*Figure 3. Specific volume scaling illustration based on PDE data. Left panel – specific volume vs. pressure measured for several temperatures. Right panel – same data where the scaled specific volume (equation 14) is plotted against the effective temperature (equation 13).*

The relationship between the two constants $P_{0,V}$ and $P_{0,T}$ is given by,

$$\frac{P_{0,V}}{P_{0,T}} = \gamma \quad (15)$$

where $\gamma$ is the Roland-Casalini power-law exponent.[56,58,60] Obviously, the scaling (13) – (14) is approximate and works only for relatively small pressures (P << $P_{0,T}$ < $P_{0,V}$); at higher pressures, the scaling is expected to break down. Even so, given that typically, $P_{0,T}$ ~ 1 *GPa*, as we will see below, the SL-TS2 scaling can be used reasonably well for pressures at least up to 200 – 300 *MPa*. The calculated SL-TS2 scaling parameters, $P_{0,T}$ and $P_{0,V}$, are summarized in Table 1 for eight glass-formers, four low-molecular-weight organic and four polymeric, based on the tabulated data from ref. [61] In calculating these parameters, we used equation 15 and the relationship,

$$\frac{dT_g}{dP} = \frac{T_g(P=0)}{P_{0,T}} \qquad (16)$$

The parameters $T_g$, $\gamma$, and $dT_g/dP$ for PDE, OTP, and PCB are regressed based on experimental data and are slightly – but within experimental error, except for PCB – different from the values published in ref. [61]

Table 1. Scaling-related parameters for the eight glass-formers. The first three columns are taken from ref. [61], except for PDE, OTP, and PCB, while the last two are calculated using Equations 15 and 16.

| Material | $T_g$, K | $\gamma$ | $dT_g/dP$, K/MPa | $P_{0,T}$, MPa | $P_{0,V}$, MPa |
|---|---|---|---|---|---|
| glycerol | 183 | 1.3 | 0.038 | 4816 | 6261 |
| PDE | 294 | 4.6 | 0.228 | 1289 | 5932 |
| OTP | 248 | 4.5 | 0.248 | 1000 | 4500 |
| PCB | 269 | 6.0 | 0.292 | 921 | 5527 |
| PCHMA | 336 | 2.1 | 0.245 | 1371 | 2880 |
| PVAc | 304 | 2.5 | 0.245 | 1241 | 3102 |
| PS | 353 | 2.1 | 0.328 | 1076 | 2260 |
| PMMA | 380 | 1.8 | 0.258 | 1473 | 2651 |

Below, we will describe the master curves for the same eight glass-formers. The model predictions need to be understood as follows – the figures represent the dielectric relaxation and specific volume vs. temperature for the ambient pressure measurements; the predictions for higher pressures can be obtained using the scaling relationships (13) and (14).

*The Parameterization of the Model*

Once the master curves are calculated, we can use them to parameterize the SL-TS2 model for each material. The parameterization process is done as follows (below, we implicitly set $P = 0$). First, we use the specific volume vs. temperature data to estimate thermodynamic parameters $T^*$, $V_{sp0}$, $a_{LL}$, $r_L$, and $r_S$; this regression is done using a Monte Carlo process combined with a trial-and-error approach for the selection of initial values. (Note that due to our earlier assumption of the Berthelot geometric mean for the cross-interaction, we automatically have $\alpha_{LS} = \sqrt{\alpha_{LL}}$). Next, we fit the dielectric $\alpha$-relaxation data by optimizing $\log(\tau_\infty)$, $E_{L,0}$, and $E_{S,0}$; this is done using Excel GRG minimization procedure. The optimization results are summarized in Table 2. Below, we compare the model fits with experimental data.

*Table 2. SL-TS2 parameters for the eight glass-formers. The parameters are regressed based on the PVT and dielectric relaxation data (see text for details). "Gly." Stands for glycerol.*

| Param. | Units | PDE | PCB | OTP | Gly. | PMMA | PS | PCHMA | PVAc |
|---|---|---|---|---|---|---|---|---|---|
| $\log(\tau_\infty)$ | | -12.0 | -16.8 | -14.5 | -17.0 | -15.9 | -14.8 | -14.0 | -15.0 |
| $E_{L,0}$ | kJ/mol | 20.0 | 55.0 | 31.0 | 53.2 | 86.5 | 69.5 | 76.3 | 37.8 |
| $E_{S,0}$ | kJ/mol | 131 | 167 | 132 | 103 | 210 | 210 | 183 | 170 |
| $T^*$ | K | 922 | 666 | 620 | 725 | 816 | 832 | 1,071 | 871 |
| $V_{sp0}$ | cm³/g | 0.696 | 0.566 | 0.839 | 0.730 | 0.793 | 0.898 | 0.867 | 0.785 |
| $\alpha_{LL}$ | | 0.9539 | 0.9956 | 0.9611 | 0.9653 | 0.9864 | 0.9798 | 0.9916 | 0.9302 |
| $r_L$ | | 466 | 3085 | 645 | 1021 | 1444 | 1091 | 3110 | 242 |
| $r_S$ | | 445 | 3110 | 631 | 986 | 1410 | 1052 | 3086 | 226 |

*Specific Volume vs. Temperature*

In Figure 4, we plot the calculated (lines) vs. measured (symbols) specific volume vs. effective temperature data for the four organic glass-formers (glycerol, ortho-terphenyl, PDE, and PCB-62). For glycerol, we used only the ambient-pressure data, while for the other three materials, multiple-pressure measurements were utilized, as described in the previous section. Note both good scaling (various datapoints collapse onto a single material-specific master curve) and fair agreement between theory and experiment. Similar to earlier papers,[23,63] here the dashed lines represent the "equilibrium" calculation and the solid lines represent the "constant cooling rate", $q$ = 1 K/min, calculation (see previous section for details).

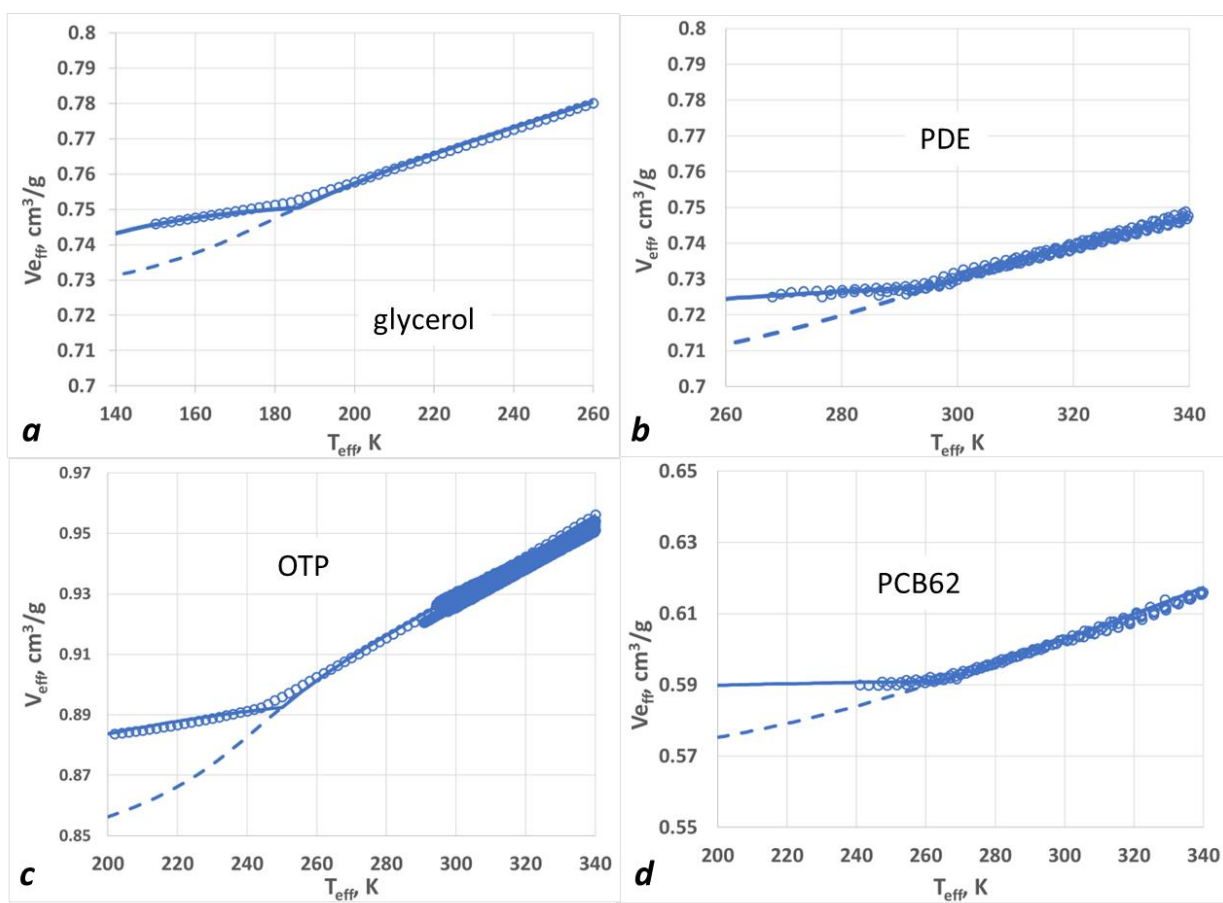

Figure 4. The "effective" specific volume (in cm³/g) vs. the "effective" temperature (in K) for four low-molecular-weight organic glass-formers: (a) glycerol; (b) PDE; (c) OTP; (d) PCB-62. In all cases, the dashed blue lines are "equilibrium" model and the solid blue lines are "constant cooling rate" model predictions.

Similar behavior can be seen for amorphous polymers (Figure 5). Here, we show only the ambient pressure data and refer the readers to earlier papers on the elevated-pressure behavior.[61] Again, to predict the specific volumes at elevated pressures, one needs to use equations (13) – (14).

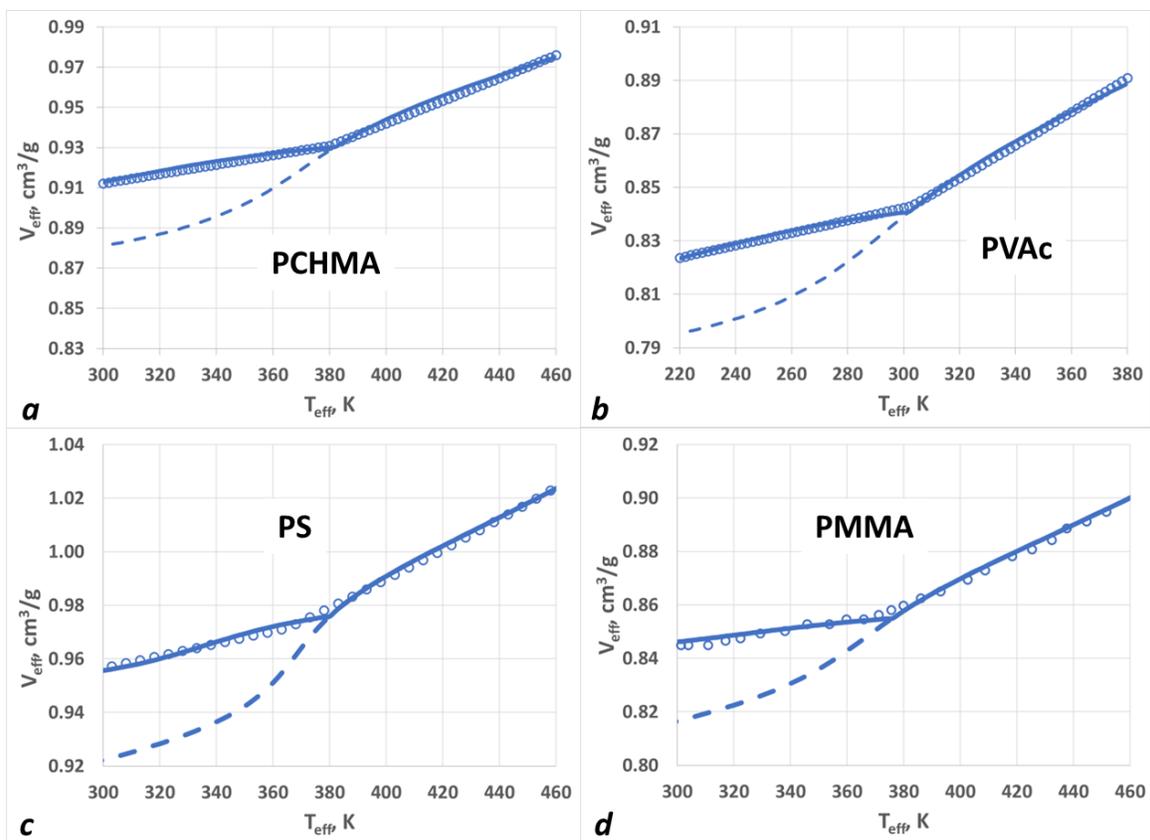

*Figure 5. Similar to Figure 4, but for polymeric glass-formers. (a) PCHMA; (b) PVAc; (c) PS; (d) PMMA. (Note that PS and PMMA results were previously published in Ref.[63])*

### Dielectric α-Relaxation vs. Temperature

In Figure 6, dielectric $\alpha$-relaxation data are plotted for the four organic glass-formers. Again, the X-axis is the "effective" temperature, rescaled to the ambient pressure case. The data for OTP, PDE, and PCB are combined from multiple experiments performed at multiple pressures. Once again, it can be seen that the data collapse well onto single master curves, well-described by the TS2 model.

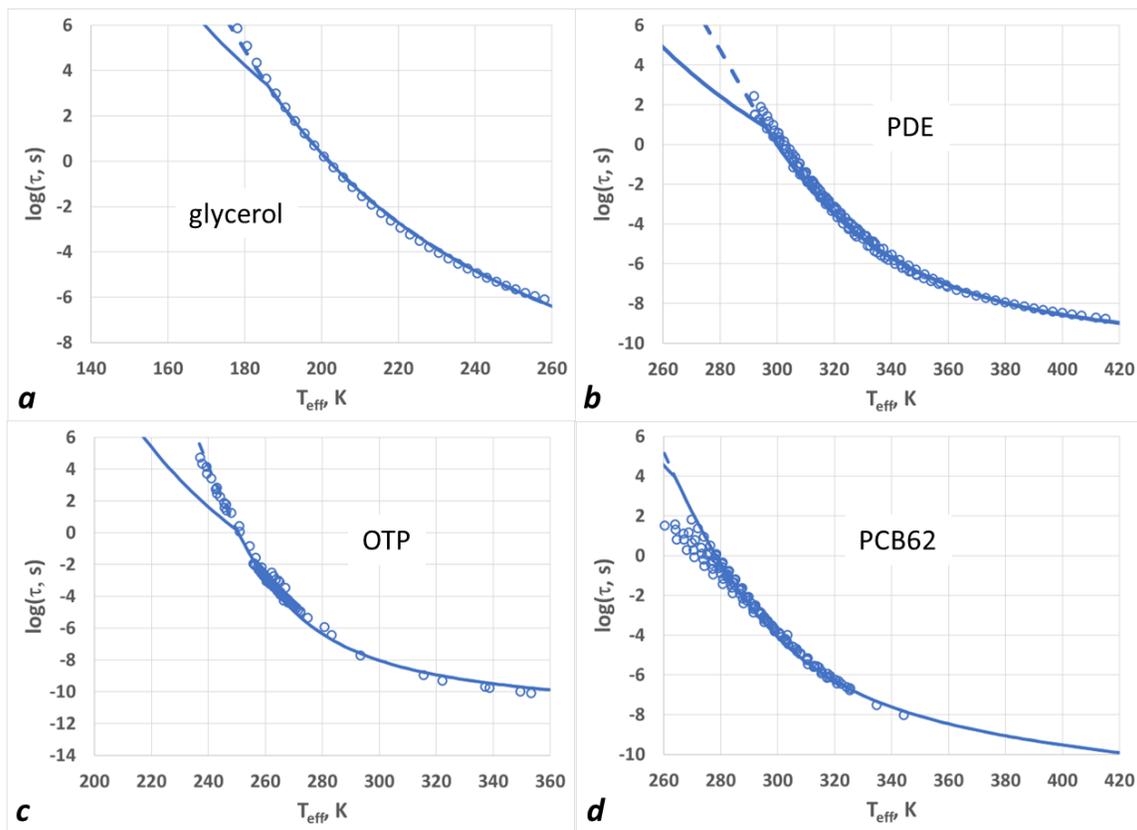

Figure 6. Logarithm of the α-relaxation time as a function of the "effective" temperature for the four organic glass-formers: (a) glycerol, (b) PDE, (c) ortho-terphenyl, and (d) PCB. The symbols are experimental data (see text for more details), the dashed lines are "equilibrium" predictions, and the solid lines are the "constant cooling rate" predictions.

Figure 7 shows the same comparison (TS2 model vs. dielectric relaxation data) for the four polymeric glass-formers (PMMA, PS, PCHMA, and PVAc). Again, we use only the ambient pressure data, but given that the τTV scaling has been shown to work for all of these polymers, we expect that the model should be able to predict the higher-pressure relaxation with the help of simple temperature rescaling (equation 13).

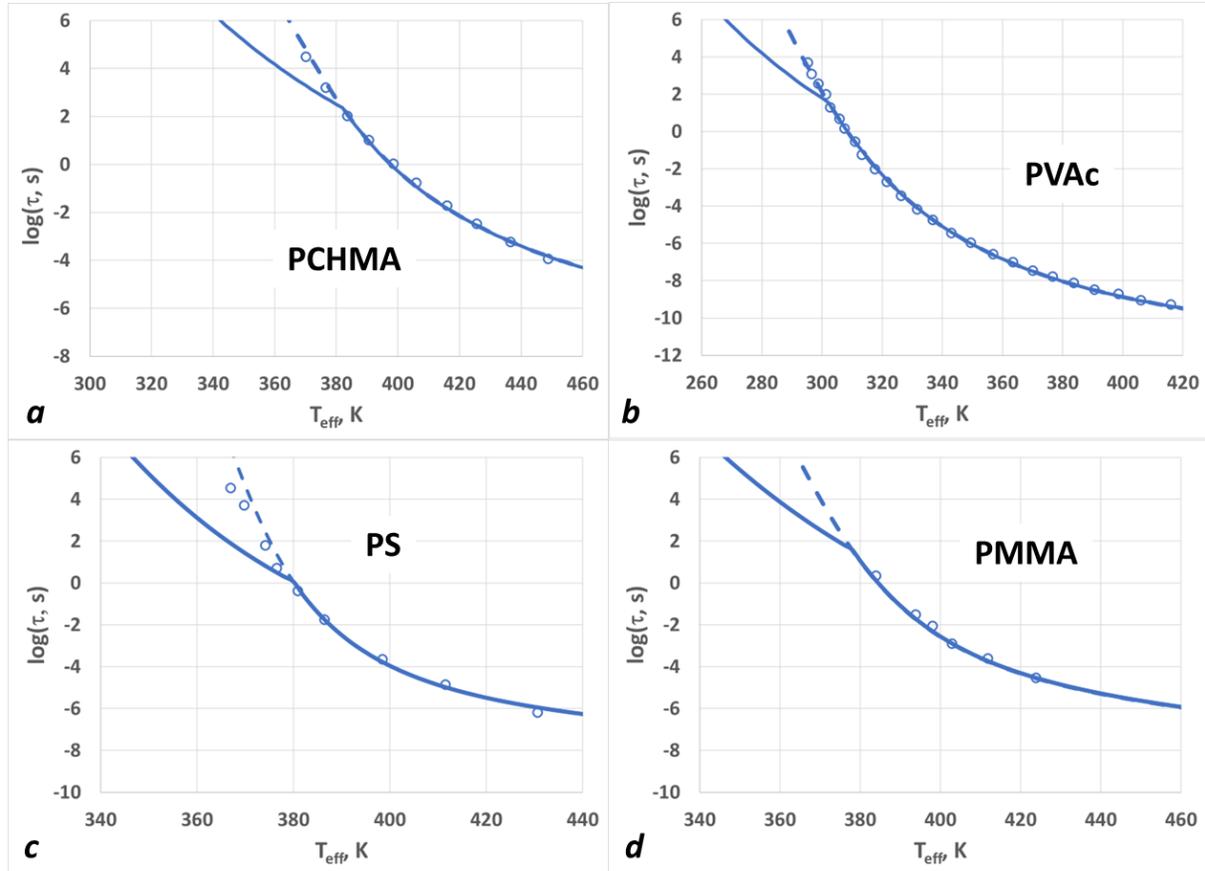

*Figure 7. Same as in Figure 6, but for the four polymeric glass-formers. (a) PCHMA, (b) PVAc, (c) PS, and (d) PMMA.*

The fact that the dielectric relaxation time data are closer to the equilibrium calculation than the specific volume data is not surprising. The specific volume data are measured at constant pressure cooling the sample at a constant cooling rate typically 0.5 – 1 K/min, while the dielectric relaxation measurements are performed waiting for long equilibration times during cooling or compression that are typically of the order or 0.5 – 1 hr. Thus, the samples have significantly more time to equilibrate during the dielectric measurements than during the specific volume measurements.

*Discussion*

We have expanded our earlier work on modeling dielectric relaxation and PVT in organic and polymeric glass-formers using SL-TS2 approach. To begin with, we showed that SL-TS2 is naturally compatible with the Casalini-Roland thermodynamical scaling, so that both density-temperature and relaxation time-temperature data collected at various pressures can be collapsed onto unique material-dependent master curves. Next, we showed that those master curves can be successfully described by the SL-TS2 equations, and summarized thermodynamic and relaxation dynamic parameters for the eight glass-formers studied here. The SL-TS2 allows the extension of the scaling in the glass state, while the Casalini-Roland thermodynamical scaling has been demonstrated only in the liquid phase.

It is important to note that the model is characterized by a relatively large number of parameters ($T_g$, $\gamma$, and $P_{0,v}$ from the scaling analysis; $T^*$, $V_{sp0}$, $a_{LL}$, $r_L$, and $r_S$ from the density-temperature data at ambient pressure; and $\log(\tau_\infty)$, $E_{L,0}$, and $E_{S,0}$ from the dielectric relaxation data at ambient pressure). While the number of parameters for the scaling and for the dielectric relaxation description is reasonable, one can pose the question whether the use of the Sanchez-Lacombe framework for thermodynamics introduces any unphysical complexity. In principle, the specific volume as a function of temperature is characterized by three constants – the specific volume at $T_g$, the coefficient of volumetric thermal expansion (CTE) of the glassy state, and the CTE of the liquid state. Describing these three constants with five parameters of the two-state Sanchez-Lacombe probably means that there are some hidden correlations between various SL parameters that we do not fully understand as yet. We plan to consider this possibility in the future, as well as investigate other possible equations of state (EoS) in place of Sanchez-Lacombe. Another way to test and potentially improve the theory and parameterization is to model the volume relaxation following the "up" and "down" temperature jumps,[69,77,80,87] which is a subject of ongoing research.

For polymeric materials, the material properties depend strongly on molecular weight, especially for relatively short polymers (M < 10,000 g/mol).[1,88–93] Here, we consider only the high-molecular weight limit (M > 100,000 g/mol). In principle, the SL-TS2 parameters for polymers should be molecular-weight dependent, in the same way as VFT or WLF parameters are. This will be another topic for future studies.

## Conclusions

We applied the SL-TS2 (two-state Sanchez-Lacombe) framework to describe PVT and dielectric $\alpha$-relaxation data for eight amorphous glass-forming materials (four organic and four polymeric). For all these materials, the dielectric and PVT data were previously shown to obey the Casalini-Roland scaling, $\tau_\alpha = f(Tv_{sp}^\gamma)$. Here, we first utilized the two-state SL-TS2 framework to successfully describe the dielectric relaxation and the volumetric thermal expansion for all these glass-formers at ambient pressure. Next, we showed that the Casalini-Roland scaling emerges naturally within SL-TS2, provided that the "units" of energy and density are all rescaled appropriately with pressure. The resulting model fits for the relaxation time and specific volume can be applied for low and moderate (up to ~200 MPa) pressures extending also in the glassy state.

## Conflicts of Interest

There are no conflicts of interest to declare.

## Acknowledgments

R.C. acknowledges the support of the Office of Naval Research (N0001421WX00833-B).

**TOC Graphics**

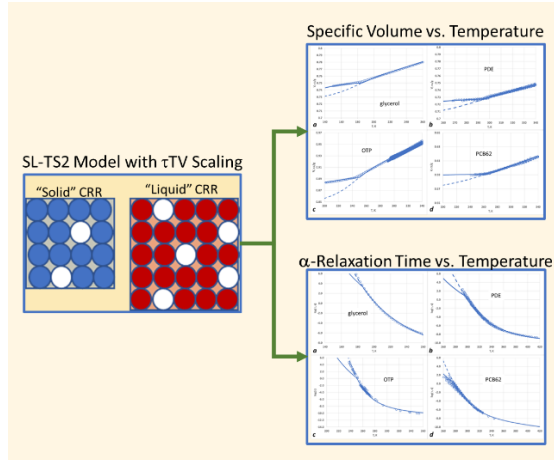